\documentclass[12pt,tightenlines,aps]{revtex4}
\newcommand{\bold}[1]{\mbox{\boldmath $#1$}}
\newcommand{\wlim}{\text{w}\!\lim}
\tighten
\begin{document}
\title{Regularization of the second-order partial derivatives
of the Coulomb potential of a point charge\footnote {This paper is
written by V Hnizdo in his private capacity. No official support or
endorsement by the Centers for Disease Control and Prevention is
intended or should be inferred.}}
\author{ V Hnizdo}
\affiliation{National Institute for Occupational Safety and Health,
Morgantown, West Virginia 26505, USA}

\begin{abstract}
\noindent{\bf Abstract}\\
The second-order partial derivatives of the Coulomb potential of a
point charge can be regularized using the Coulomb potential of a
charge of the oblate spheroidal shape that a moving
rest-frame-spherical charge acquires by the Lorentz contraction.
This `physical' regularization is shown to be fully equivalent to
the standard delta-function identity involving these derivatives.
\end{abstract}
\maketitle

\vspace{5ex} \noindent Quantities with a singularity of the type
$1/r^3$ at the origin $r=0$ occur in classical electrodynamics in
connection with the idealization of a point charge distribution. For
example, the straightforward calculation of the second-order partial
derivatives of the Coulomb potential $1/r$ of a unit point charge
yields $\partial^2 r^{-1}/\partial x_i\partial
x_j=(3x_ix_j-r^2\delta_{ij})/r^5$, or the field of a point electric
or magnetic dipole, obtained as the straightforward gradient of a
potential with radial dependence $1/r^2$, has the radial dependence
$1/r^3$. Because of the $1/r^3$ singularity at the origin, the
integral of such a quantity over any three-dimensional region that
includes the origin $r=0$ does not exist even in the
improper-integral sense: the value of the integral obtained by
excluding from the integration a region ${\cal V}_0$ around the
origin and taking the limit of the size of ${\cal V}_0$ tending to
zero depends on the shape and orientation of ${\cal V}_0$. Integrals
involving derivatives of $1/r^2$ or second-order derivatives of
$1/r$ therefore have to be suitably regularized. A formal way of
doing that is to use the delta-function identity \cite{Frahm}
\begin{equation}
\frac{\partial^2}{\partial x_i\partial x_j}\frac{1}{r}
=\frac{3x_ix_j-r^2\delta_{ij}}{r^5}-\frac{4\pi}{3}\delta_{ij}\delta(\bold{ r})
\label{laplace/3}
\end{equation}
where $\delta_{ij}$ is the Kronecker delta
symbol and $\delta(\bold{ r})=\delta(x_1)\delta(x_2)\delta(x_3)$ is the
three-dimensional delta function. The validity of the identity (\ref{laplace/3})
can be justified most easily by the use of the straightforward regularization
$1/(r^2+a^2)^{1/2}$ of the singular potential $1/r$:
\begin{equation}
\lim_{a\rightarrow 0}\frac{\partial^2}{\partial x_i\partial x_j}
\frac{1}{\sqrt{r^2+a^2}}
=\lim_{a\rightarrow 0}\frac{3x_ix_j-r^2\delta_{ij}}{(r^2+a^2)^{5/2}}
-\lim_{a\rightarrow 0}\frac{a^2\delta_{ij}}{(r^2+a^2)^{5/2}}
\label{reg}
\end{equation}
since here the second term on the right-hand side is a well-known
representation of $-\case{4}{3}\pi\delta_{ij}\delta(\bold{ r})$
(e.g., see \cite{Jack}). The first term on the right-hand side of
the identity (\ref{laplace/3}) is as such still non-integrable at
the origin $r=0$, but the regularization (\ref{reg}) also includes a
specification of the regularization of this term; of course, the
limits $a\rightarrow 0$ are understood to be taken only after a
three-dimensional integration with a well-behaved `test' function.
Regularizing the term $(3x_ix_j{-}r^2\delta_{ij})/r^5$ in a
different, but equivalent, way, the identity (\ref{laplace/3}) may
be written as
\begin{equation}
\frac{\partial^2}{\partial x_i\partial x_j}\frac{1}{r}
=\lim_{a\rightarrow 0+}\frac{3x_ix_j-r^2\delta_{ij}}{r^5}\,
\Theta(r-a)-\frac{4\pi}{3}\delta_{ij}\delta(\bold{ r})
\label{laplace/3Theta}
\end{equation}
where $\Theta(\cdot)$ is the Heaviside step function. This formulation is
equivalent to the stipulation that the spherical coordinates are to be used
in the integration with a test function and that the angular integration
is to be done first; the identity (\ref{laplace/3}) has been derived in
reference \cite{Frahm} effectively in the form (\ref{laplace/3Theta}).

In a recent paper on the Coulomb-gauge vector potential of a
uniformly moving point charge \cite{H}, an occasion has arisen of
using the delta-function identity (\ref{laplace/3Theta}) for the
regularization of an integral of the type $\int {\rm
d}^3r\,f(\bold{r})\partial^2r^{-1}/\partial x_i^2$ in terms of which
the difference between the Coulomb- and Lorenz-gauge vector
potentials in that problem can be obtained as the solution to a
Poisson equation. Since the relation (\ref{laplace/3Theta}) is an
{\it identity}, there should be no doubt as to the correctness of
such a formal regularization. However, in a problem that concerns a
moving charge, it would be reassuring if one could show that a more
`physical' regularization procedure will yield the same results.
Physically, it is natural to regularize the Coulomb potential $1/r$
of a moving point charge by the Coulomb potential of a charge that
has the oblate spheroidal shape that a moving rest-frame-spherical
`elementary' charge of finite extension $a$ acquires by the Lorentz
contraction, and then to take the limit $a\rightarrow 0$ of any
integral involving second-order derivatives of this potential. Such
regularization involves an `ellipsoidal' approach toward the
singularity at the origin because of the spheroidal shape of the
moving elementary charge---in contrast to the first term on the
right-hand side of identity (\ref{laplace/3Theta}) that stipulates a
strictly `spherical' approach toward this singularity. In this note,
we demonstrate that a physical regularization along the above lines
is indeed fully equivalent to the delta-function identity
(\ref{laplace/3Theta}). We believe that the reward for carrying out
the calculation that this demonstration requires will be a physical
insight into and ensuing confidence in use of a useful formal
relation.

As a preliminary, we note that a `physical' justification of the well-known
delta-function identity $\nabla^2(1/r)=-4\pi\delta(\bold{r})$ can be provided
very simply. Let $\varphi_a(\bold{r})$ be the Coulomb potential of
a unit elementary charge described by the density
$\rho_a(\bold{r})=(1/V_a)\Theta(a^2{-}\gamma^2x_1^2{-}x_2^2{-}x_3^2)$, where
$V_a=\case{4}{3}\pi a^3/\gamma$,  which is the density of a uniformly
charged spheroid with semiaxes $a/\gamma$, $a$, $a$, centred at the origin.
For $\gamma=(1-v^2/c^2)^{-1/2}$, this is the density of a charge that is
moving with a velocity $v$ ($c$ is the speed of light) along
the $x_1$-axis and that is a uniform ball of radius $a$ in its rest frame.
Then, for any well-behaved test function $f(\bold{r})$, we have
by the fact that the potential $\varphi_a(\bold{r})$
satisfies the Poisson equation
$\nabla^2\varphi_a(\bold{r})=-4\pi\rho_a(\bold{r})$:
\begin{equation}
\int {\rm d}^3r\,f(\bold{r})\nabla^2\varphi_a(\bold{r})
=-4\pi\int {\rm d}^3r\,f(\bold{r})\rho_a(\bold{r})
=-\frac{4\pi}{V_a}\int_{{\cal V}_a}{\rm d}^3r\,f(\bold{r})
=-\frac{4\pi}{V_a}V_a f(\bold{r}_0)
\end{equation}
where the mean-value theorem is used on the right-hand side, with
$\bold{r}_0$ being a point inside the region ${\cal V}_a$ occupied
by the spheroid. Taking now the limit $a\rightarrow 0$, we obtain
\begin{equation}
\lim_{a\rightarrow 0}
\int {\rm d}^3r\,f(\bold{r})\nabla^2\varphi_a(\bold{r})
=-4\pi\lim_{a\rightarrow 0}f(\bold{r}_0)
=-4\pi f(0)=-4\pi\int {\rm d}^3r f(\bold{r})\delta(\bold{r})
\end{equation}
because the point $\bold{r}_0\in{\cal V}_a$ has to converge on the
origin $\bold{r}=0$ as $a\rightarrow 0$, and thus we can write
\begin{equation}
\lim_{a\rightarrow 0}\nabla^2\varphi_a(\bold{r})=-4\pi\delta(\bold{r}).
\end{equation}

A uniformly charged ball was not the most popular model of an
elementary charge employed in the classical electron theory---this
was a uniformly charged spherical shell, or, equivalently,  a
charged spherical conductor (see, e.g., \cite{Miller}). The Coulomb
potential of a uniformly moving charged conductor that is spherical
with radius $a$ in its rest frame is the most convenient one to use
for our purpose because it equals the Coulomb potential of a charged
conducting oblate spheroid of semiaxes $a/\gamma$, $a$, $a$ (see
\cite{TGMPF}; an interesting historical background to the problem of
a moving charged sphere can be found in \cite{Red}), and this
potential can be expressed in terms of an elementary function
\cite{LL}:
\begin{equation}
\varphi_a(\bold{r})=\left\{\begin{array}{ll}
(1/\beta a)\arctan[\beta a/\sqrt{\case{1}{4}(r_+{+}r_-)^2{-}(\beta a)^2}]
&\text{ for }\;\gamma^2x_1^2+\rho^2\ge a^2\\
(1/\beta a)\arctan(\beta\gamma) &\text{ for }\;\gamma^2x_1^2+\rho^2<a^2
\end{array}\right.
\label{phia}
\end{equation}
where
\begin{equation}
r_{\pm}=\sqrt{x_1^2+(\rho\pm\beta a)^2}\;\;\;\;\;\;\rho=\sqrt{x_2^2+x_3^2}
\;\;\;\;\;\;\gamma=\frac{1}{\sqrt{1-\beta^2}}\;\;\;\;\;\;\beta=\frac{v}{c}.
\label{rr}
\end{equation}
The partial derivatives
$\partial\varphi_a(\bold{r})/\partial x_i$ of the
potential $\varphi_a(\bold{r})$ for $\gamma^2x_1^2+\rho^2\ge a^2$ are
\begin{eqnarray}
\frac{\partial\varphi_a(\bold{r})}{\partial x_1}
\bigg|_{\gamma^2x_1^2+\rho^2\ge a^2}
&=&-\frac{x_1}{r_+r_-\sqrt{\case{1}{4}(r_++r_-)^2-(\beta a)^2}}
\label{firstd1}
\\
\frac{\partial\varphi_a(\bold{r})}{\partial x_{2,3}}
\bigg|_{\gamma^2x_1^2+\rho^2\ge a^2}
&=&-\frac{(r_- -r_+)\beta a+(r_+ +r_-)\rho}
{r_+ r_-\sqrt{\case{1}{4}(r_+ +r_-)^2-(\beta a)^2}}
\,\frac{x_{2,3}}{(r_+ +r_-)\rho}.
\label{firstd23}
\end{eqnarray}
Since the potential $\varphi_a(\bold{r})$ is constant inside the spheroid
$\gamma^2x_1^2+\rho^2=a^2$, the partial derivatives
$\partial\varphi_a(\bold{r})/\partial x_i$ can be written as
$\Theta[(\gamma^2x_1^2{+}\rho^2)^{1/2}{-}a]
\partial\varphi_a(\bold{r})/\partial x_i$,
and thus the second-order partial derivatives
$\partial^2\varphi_a(\bold{r})/\partial x_i\partial x_j$
can be written as
\begin{eqnarray}
\frac{\partial^2\varphi_a(\bold{r})}{\partial x_i\partial x_j}
&&=\frac{\partial^2\varphi_a(\bold{r})}{\partial x_i\partial x_j}\,
\Theta(\sqrt{\gamma^2x_1^2{+}\rho^2}{-}a)
+\frac{\partial\varphi_a(\bold{r})}{\partial x_i}
\frac{\partial}{\partial x_j}\Theta(\sqrt{\gamma^2x_1^2{+}\rho^2}{-}a)
\nonumber \\
&&=
\frac{\partial^2\varphi_a(\bold{r})}{\partial x_i\partial x_j}\,
\Theta(\sqrt{\gamma^2x_1^2{+}\rho^2}{-}a)
+\frac{\partial\varphi_a(\bold{r})}{\partial x_i}\,
\frac{[1{+}(\gamma^2{-}1)\delta_{1j}]x_j}{\sqrt{\gamma^2x_1^2+\rho^2}}\,
\delta(\sqrt{\gamma^2x_1^2{+}\rho^2}{-}a)
\label{dervs}
\end{eqnarray}
where the derivatives of the potential on the right-hand
side are understood as those of the expression for
the potential exterior to the spheroid.

The equivalence of the delta-function identity (\ref{laplace/3Theta})
and the regularization that uses the Coulomb
potential $\varphi_a(\bold{r})$ of a charged conducting spheroid
demands that, for any well-behaved test function $f(\bold{r})$,
\begin{equation}
\lim_{a\rightarrow 0}\int{\rm d}^3r\,f(\bold{r})
\frac{\partial^2\varphi_a(\bold{r})}{\partial x_i\partial x_j}
=\lim_{a\rightarrow 0+}\int_{r> a}{\rm d}^3r\,f(\bold{r})
\frac{3x_ix_j-r^2\delta_{ij}}{r^5} -\frac{4\pi}{3}\delta_{ij}f(0).
\label{equiv}
\end{equation}
Using (\ref{dervs}), we write the left-hand side of (\ref{equiv}) as
\begin{eqnarray}
\lim_{a\rightarrow 0}\int{\rm d}^3r\,f(\bold{r})
\frac{\partial^2\varphi_a(\bold{r})}{\partial x_i\partial x_j}
=&&\!\!\!\!\!\lim_{a\rightarrow 0}\int_{\gamma^2x_1^2+\rho^2> a^2}
{\rm d}^3r\,f(\bold{r})
\,\frac{\partial^2\varphi_a(\bold{r})}{\partial x_i\partial x_j}
\nonumber \\
&&\!\!\!\!\!+[1{+}(\gamma^2{-}1)\delta_{1j}]\lim_{a\rightarrow 0}
\int{\rm d}^3r\,f(\bold{r})\,
\frac{\partial\varphi_a(\bold{r})}{\partial x_i}\,
\frac{x_j\delta(\sqrt{\gamma^2x_1^2{+}\rho^2}{-}a)}
{\sqrt{\gamma^2x_1^2+\rho^2}}. \label{lhs}
\end{eqnarray}
Using the expressions (\ref{firstd1}) and (\ref{firstd23}) for the
derivatives $\partial\varphi_a(\bold{r})/\partial x_i$,
the second term on the right-hand side
of (\ref{lhs}) can be evaluated in closed form.
Let us first assume that $i{=}j{=}1$. Transforming as
$\gamma x_1\rightarrow x_1$ and then
to the spherical coordinates $r$, $\theta$, $\phi$,
with $x_1$ as the polar axis and $\cos\theta=\xi$, we obtain
\begin{eqnarray}
&&\gamma^2\lim_{a\rightarrow 0}
\int{\rm d}^3r\,f(\bold{r})\,
\frac{\partial\varphi_a(\bold{r})}{\partial x_1}\,
\frac{x_1\delta(\sqrt{\gamma^2x_1^2+\rho^2}-a)}
{\sqrt{\gamma^2x_1^2+\rho^2}}\nonumber \\
&&\;\;\;=\lim_{a\rightarrow 0}\int {\rm d}^3r\,
f(x_1/\gamma,x_2,x_3)\,
\frac{\partial\varphi_a(\bold{r})}{\partial x_1}
\bigg|_{x_1\rightarrow x_1/\gamma}
\frac{x_1\delta(r-a)}{r}\nonumber \\
&&\;\;\;=-\lim_{a\rightarrow 0}
\int_{-1}^1{\rm d}\xi\int_0^{2\pi}{\rm d}\phi
\,f(x_1/\gamma,x_2,x_3)|_{r=a}\,\frac{\xi^2}{1-\beta^2(1-\xi^2)}
\nonumber \\
&&\;\;\;=-2\pi f(0)\int_{-1}^{1}\frac{{\rm d}\xi\,\xi^2}
{1-\beta^2(1-\xi^2)}
=-2\pi\left(\frac{2}{\beta^2}-\frac{2\arcsin\beta}
{\gamma\beta^3}\right)f(0)
\label{dphi1}
\end{eqnarray}
where
$\partial\varphi_a(\bold{r})/\partial x_1|_{x_1\rightarrow x_1/\gamma}$
denotes the exterior partial derivative (\ref{firstd1}) after the
transformation $\gamma x_1\rightarrow x_1$.
Here, the delta function $\delta(r-a)$ led to
an immediate radial integration, which enabled a considerable
simplification of the integrand;
the limit $a\rightarrow 0$ then could be taken inside
the remaining integral, yielding
\begin{eqnarray}
\lim_{a\rightarrow 0}f(x_1/\gamma,x_2,x_3)|_{r=a}
&&=\lim_{a\rightarrow 0}
f(a\cos\theta/\gamma,a\cos\phi\sin\theta,a\sin\phi\sin\theta)
\nonumber \\
&&=f(0,0,0)\equiv f(0).
\end{eqnarray}
A similar calculation for $i{=}j{=}2$ yields
\begin{eqnarray}
&&\lim_{a\rightarrow 0}\int {\rm d}^3r\,f(\bold{r})\,
\frac{\partial\varphi_a(\bold{r})}{\partial x_2}\,
\frac{x_2\delta(\sqrt{\gamma^2x_1^2+\rho^2}-a)}
{\sqrt{\gamma^2x_1^2+\rho^2}} \nonumber \\
&&\;\;\;=\lim_{a\rightarrow 0}\int\frac{{\rm d}^3r}{\gamma}\,
f(x_1/\gamma,x_2,x_3)\frac{\partial\varphi_a(\bold{r})}{\partial x_2}
\bigg|_{x_1\rightarrow x_1/\gamma}\frac{x_2\delta(r-a)}{r}
\nonumber \\
&&\;\;\;=-\frac{f(0)}{\gamma^2}
\int_{-1}^1{\rm d}\xi\int_0^{2\pi}{\rm d}\phi
\,\frac{(1-\xi^2)\cos^2\phi}{1-\beta^2(1-\xi^2)}
\nonumber \\
&&\;\;\;=-\frac{\pi f(0)}{\gamma^2}
\int_{-1}^1{\rm d}\xi\,\frac{1-\xi^2}{1-\beta^2(1-\xi^2)}
=-2\pi\left(1-\frac{1}{\beta^2}+\frac{\arcsin\beta}
{\gamma\beta^3}\right)f(0).
\label{dphi2}
\end{eqnarray}
Here, similarly as in the integration in (\ref{dphi1}),
even with the relatively complicated expression
(\ref{firstd23}) for the exterior partial derivative
$\partial\varphi_a(\bold{r})/\partial x_2$,
the integrand could be simplified considerably after the radial
integration.  The case $i{=}j{=}3$ will obviously yield the same
result, while for any mixed case $i\ne j$, the
integration with respect to the azimuthal angle $\phi$ will
lead to a vanishing result. Collecting these results, we have
\begin{equation}
[1{+}(\gamma^2{-}1)\delta_{1j}]
\lim_{a\rightarrow 0}\int {\rm d}^3r\,f(\bold{r})\,
\frac{\partial\varphi_a(\bold{r})}{\partial x_i}\,
\frac{x_j\delta(\sqrt{\gamma^2x_1^2{+}\rho^2}{-}a)}
{\sqrt{\gamma^2x_1^2+\rho^2}}=-2\pi g_{ij}(\beta)f(0)
\label{summary}
\end{equation}
where
\begin{equation}
g_{ij}(\beta)=\left\{\begin{array}{ll}
2/\beta^2-(2/\gamma\beta^3)\arcsin\beta & \text{ for }
i{=}j{=}1 \\
1-1/\beta^2+(1/\gamma\beta^3)
\arcsin\beta & \text{ for } i{=}j{=}2, 3\\0 & \text{ for } i\ne j.
\end{array}\right.
\end{equation}
We note that $\lim_{\beta\rightarrow
0}g_{ij}(\beta)=\case{2}{3}\delta_{ij}$. Using  (\ref{lhs}) and
(\ref{summary}), the condition (\ref{equiv}) of regularization
equivalence can now be written as
\begin{eqnarray}
&&\lim_{a\rightarrow 0}\int_{\gamma^2x_1^2+\rho^2> a^2} {\rm
d}^3r\,f(\bold{r}) \,\frac{\partial^2\varphi_a(\bold{r})}{\partial
x_i\partial x_j} -\lim_{a\rightarrow 0+}\int_{r> a}{\rm
d}^3r\,f(\bold{r})
\frac{3x_ix_j-r^2\delta_{ij}}{r^5}\nonumber \\
&&\;\;\;=2\pi[g_{ij}(\beta)-\textstyle{\frac{2}{3}}\delta_{ij}]f(0).
\label{equiv2}
\end{eqnarray}

To prove that the condition (\ref{equiv2}) holds true,
we proceed as follows. The first limit on the left-hand side of
(\ref{equiv2}) can be written  more simply as
\begin{equation}
\lim_{a\rightarrow 0}\int_{\gamma^2x_1^2+\rho^2> a^2} {\rm
d}^3r\,f(\bold{r}) \,\frac{\partial^2\varphi_a(\bold{r})}{\partial
x_i\partial x_j} =\lim_{a\rightarrow 0}\int_{\gamma^2x_1^2+\rho^2>
a^2} {\rm d}^3r\,f(\bold{r})\frac{3x_ix_j-r^2\delta_{ij}}{r^5}
\label{2ndder}
\end{equation}
since the derivatives $\partial^2\varphi_a(\bold{r})/\partial
x_i\partial x_j$ for $\gamma^2x_1^2+\rho^2>a^2$ can be expanded in
powers of $(\beta a)^2$ (such an expansion is obtained most easily
by differentiating term by term the corresponding expansion of
$\varphi_a$), where the zeroth-order term equals
$(3x_ix_j-r^2\delta_{ij})/r^5$ and the integrals involving the
higher-order terms vanish in the limit $a\rightarrow 0$. It suffices
to show this only for the case $i{=}j{=}1$. Here we have
\begin{equation}
\frac{\partial^2\varphi_a(\bold{r})}
{\partial x_1^2}\bigg|_{\gamma^2x_1^2+\rho^2>a^2}
=\sum_{n=0}^{\infty}
(-1)^n(2n+2)P_{2n+2}(x_1/r)\frac{(\beta a)^{2n}}{r^{2n+3}},
\label{expanphi}
\end{equation}
where $P_m(\cdot)$ are the Legendre polynomials.
Using (\ref{expanphi}) and the  multipole expansion of the function
$f(\bold{r})$ after the transformation $x_1\rightarrow x_1/\gamma$,
\begin{equation}
f(r\cos\theta/\gamma,r\cos\phi\sin\theta,r\sin\phi\sin\theta)
=\sum_{lm}f_{lm}(r,\gamma)Y_{lm}(\theta,\phi)
\label{expanf}
\end{equation}
we obtain
\begin{equation}
\int_{\gamma^2x_1^2+\rho^2> a^2}{\rm d}^3r\,
f(\bold{r})\frac{\partial^2\varphi_a(\bold{r})}
{\partial x_1^2}
=\sum_{n=0}^{\infty}c_n(a,\gamma)(\beta a)^{2n}
\label{intexpan}
\end{equation}
where
\begin{eqnarray}
c_n(a,\gamma)&=&
\sum_{lm}C_{nlm}(\gamma)\int_a^{\infty}{\rm d}r\,
\frac{f_{lm}(r,\gamma)}{r^{2n+1}}\\
C_{nlm}(\gamma)&=&(-1)^n(2n+2)\int{\rm d}\Omega\,
Y_{lm}(\theta,\phi)\frac{P_{2n+2}(\cos\theta/\gamma u)}{\gamma u^{2n+3}}
\;\;\;\;\;\;u=\sqrt{1-\beta^2\cos^2\theta}.
\label{cn}
\end{eqnarray}
We note that $C_{n00}(\gamma)=0$ for any $n\ge 0$.
As $\lim_{r\rightarrow 0}f_{lm}(r,\gamma)=0$
for any $l>0$, we have that $\lim_{a\rightarrow 0}[a^{2n}
\int_a^{\infty}{\rm d}r\,
f_{lm}(r,\gamma)/r^{2n+1}]=0$ when both $n>0$ and $l>0$, and thus
$\lim_{a\rightarrow 0}[c_n(a,\gamma)(\beta a)^{2n}]=0$ for any $n>0$.
Therefore, we indeed obtain
\begin{equation}
\lim_{a\rightarrow 0}\int_{\gamma^2x_1^2+\rho^2>a^2}{\rm d}^3r\,
f(\bold{r})\frac{\partial^2\varphi_a(\bold{r})}{\partial x_1^2}
=\lim_{a\rightarrow 0}c_0(a,\gamma)
=\lim_{a\rightarrow 0}\int_{\gamma^2x_1^2+\rho^2>a^2}{\rm d}^3r\,f(\bold{r})
\frac{3x_1^2-r^2}{r^5}.
\label{limc0}
\end{equation}
Using (\ref{2ndder}), the condition (\ref{equiv2}) can be expressed as
\begin{equation}
\lim_{a\rightarrow 0}\int_{{\cal U}_a}{\rm d}^3r\,f(\bold{r})
\frac{3x_ix_j-r^2\delta_{ij}}{r^5}
=2\pi[g_{ij}(\beta)-\textstyle{\frac{2}{3}}\delta_{ij}]f(0)
\label{intVa}
\end{equation}
where the integration region ${\cal U}_a$ is the region
between the surfaces of the oblate spheroid $\gamma^2 x_1^2+\rho^2=a^2$
and the sphere $x_1^2+\rho^2=a^2$:
\begin{equation}
{\cal U}_a =\{(x_1,x_2,x_3);\,\gamma^2x_1^2+x_2^2+x_3^2>a^2,\,
x_1^2+x_2^2+x_3^2<a^2\}. \label{Va}
\end{equation}

We now evaluate the left-hand side of (\ref{intVa}).
When the size parameter $a$ tends to zero, the integration region ${\cal U}_a$
gets progressively smaller and closer to the origin $r=0$, and thus,
for $\bold{r}\in {\cal U}_a$,  $f(\bold{r})\rightarrow f(0)$ as
$a\rightarrow 0$. We can therefore write the left-hand side of (\ref{intVa})
as
\begin{equation}
\lim_{a\rightarrow 0}\int_{{\cal U}_a}{\rm d}^3r\,f(\bold{r})
\frac{3x_ix_j-r^2\delta_{ij}}{r^5}
=f(0)\lim_{a\rightarrow 0}
\int_{{\cal U}_a}{\rm d}^3r\,\frac{3x_ix_j-r^2\delta_{ij}}{r^5}.
\label{limVa}
\end{equation}
Transforming here the integral on the right-hand side
to the spherical coordinates, with $x_1$ as the polar axis and
$\cos\theta=\xi$, we obtain for $i{=}j{=}1$:
\begin{eqnarray}
\int_{{\cal U}_a}{\rm d}^3r\,\frac{3x_1^2-r^2}{r^5}
&&=2\pi\int_{-1}^1{\rm d}\xi\,(3\xi^2-1)
\int_{a/\sqrt{1+(\gamma^2-1)\xi^2}}^a\frac{{\rm d}r}{r}
\nonumber \\
&&=\pi\int_{-1}^1{\rm d}\xi\,(3\xi^2-1)\ln[1+(\gamma^2-1)\xi^2]
\nonumber \\
&&=2\pi\left(\frac{2}{\beta^2}-
\frac{2\arcsin \beta}{\gamma\beta^3}-\frac{2}{3}\right).
\label{intVa1}
\end{eqnarray}
The case $i{=}j{=}2$ gives
\begin{eqnarray}
\int_{{\cal U}_a}{\rm d}^3r\,\frac{3x_2^2-r^2}{r^5}
&&=\frac{1}{2}\int_{-1}^1{\rm d}\xi\int_0^{2\pi}{\rm d}\phi\,
[3(1-\xi^2)\cos^2\phi-1]\ln[1+(\gamma^2-1)\xi^2] \nonumber \\
&&=\frac{\pi}{2}\int_{-1}^1{\rm d}\xi\,(1-3\xi^2)\ln[1+(\gamma^2-1)\xi^2]
\nonumber \\
&&=2\pi\left(\frac{1}{3}-\frac{1}{\beta^2}+
\frac{\arcsin \beta}{\gamma\beta^3}\right)
\label{intVa2}
\end{eqnarray}
and the same result will obviously be obtained for $i{=}j{=}3$. The mixed
cases $i\ne j$ will all yield zero on account of the integration with respect
to $\phi$. The values of the integrals (\ref{intVa1}) and (\ref{intVa2})
are independent of $a$, and using these results and (\ref{limVa}),
we obtain (\ref{intVa}).
This completes the proof of the regularization equivalence (\ref{equiv}).

In closing, we would like to stress that the regularization
equivalence (\ref{equiv}) is bound to hold also when the potential
$\varphi_a(\bold{r})$ is the Coulomb potential of a uniformly
charged spheroid, or of any other Lorentz-contracted charge
distribution that is, in its rest frame, spherically symmetric and
characterized by a finite size parameter $a$. (Explicit expressions
for the potential of a uniformly charged spheroid can be found in
the literature \cite{Muratov,Wang,Miloh}, but they are rather more
complicated than the conducting-spheroid expression (\ref{phia}).)
In fact, using the powerful results  of the theory of generalized
functions and derivatives, one can very easily give a formal proof
that (\ref{equiv}) holds for the Coulomb potential
$\varphi_a(\bold{r})$ of a charge distribution $\rho_a(\bold{r})$ of
any shape, subject only to the condition $\lim_{a\rightarrow
0}\rho_a(\bold{r})=\delta(\bold{r})$; this proof is given in
Appendix.
\newpage
\section*{Appendix}
\noindent {\it Lemma.} Let
\begin{equation}
\varphi_a(\bold{ r})=\int {\rm d}^3r'\,\frac{\rho_a(\bold{
r}')}{|\bold{r}-\bold{r}'|},
\end{equation}
where the integration extends over all space and $\rho_a(\bold{ r})$
is a localized (not necessarily spherically symmetric) function of
$\bold{r}=(x_1,x_2,x_3)$  that depends on a parameter $a$ so that
\begin{equation}
\wlim_{a\to 0}\rho_a(\bold{ r}) =\delta(\bold{ r}).
\end{equation}
Then
\begin{equation}
\wlim_{a\to 0}\frac{\partial^2\varphi_a(\bold{ r})}{\partial
x_i\partial x_j}= -\frac{4\pi}{3}\delta_{ij}\delta(\bold{
r})+\wlim_{\varepsilon\to
0+}\frac{3x_ix_j-r^2\delta_{ij}}{r^5}\,\Theta(r-\varepsilon),
\end{equation}
where $r=|\bold{ r}|=(x_1^2+x_2^2+x_3^2)^{1/2}$ and $\Theta(\cdot)$
is the Heaviside step function. The symbol $\wlim$ denotes the weak
limit: $\wlim_{a\to a_0}f_a(\bold{ r})= g(\bold{ r})$ iff
$\lim_{a\to a_0}\int{\rm d}^3r\, t(\bold{ r})f_a(\bold{ r}) =\int
{\rm d}^3r\,t(\bold{ r})g(\bold{ r})$ for any `well-behaved' test
function $t(\bold{ r})$.
\newline
{\it Proof.} To prove (34), we need to show that, for any
`well-behaved' test function $f(\bold{r})$,
\begin{equation}
\lim_{a\to 0}\int {\rm d}^3r\, f(\bold{
r})\frac{\partial^2\varphi_a(\bold{ r})}{\partial x_i\partial
x_j}=-\frac{4\pi}{3}f(0)\delta_{ij}+\lim_{\varepsilon\to
0+}\int_{r>\varepsilon} {\rm d}^3r \,f(\bold{
r})\,\frac{3x_ix_j-r^2\delta_{ij}}{r^5}.
\end{equation}
To evaluate the left-hand side of (35), we note that we can replace
the derivative $\partial^2/\partial x_i\partial x_j$ by the
generalized (distributional) derivative $\bar{\partial}^2/\partial
x_i\partial x_j$. This allows us to exchange the order of the limit
$a\to 0$ and the differentiation [11, p 12] since the space of the
generalized functions is complete [12]:
\begin{eqnarray}
\lim_{a\to 0}\int {\rm d}^3r\, f(\bold{
r})\frac{\partial^2\varphi_a(\bold{ r})}{\partial x_i\partial
x_j}&=&\lim_{a\to 0}\int {\rm d}^3r\, f(\bold{
r})\frac{\bar{\partial}^2\varphi_a(\bold{ r})}{\partial x_i\partial
x_j}
\nonumber \\
&=&\int {\rm d}^3r\, f(\bold{ r})\frac{\bar{\partial}^2}{\partial
x_i\partial x_j}\lim_{a\to
0}\varphi_a(\bold{ r})\nonumber\\
&=&\int {\rm d}^3r \,f(\bold{ r})\frac{\bar{\partial}^2}{\partial
x_i\partial x_j}\frac{1}{r}.
\end{eqnarray}
Here, the 3rd line was obtained using (32) and (33). But
\begin{equation}
\frac{\bar{\partial}^2}{\partial x_i\partial x_j}\frac{1}{r}=
-\frac{4\pi}{3}\delta_{ij} \delta(\bold{ r})+\wlim_{\varepsilon\to
0+}\frac{3x_ix_j-r^2\delta_{ij}}{r^5}\,\Theta(r-\varepsilon)
\end{equation}
(see [11], p 28, but note that the signs of the right-hand sides of
(3.129) and (3.130) are there misprinted; see also [13,14]), and
thus
\begin{equation}
\int {\rm d}^3r\, f(\bold{ r})\frac{\bar{\partial}^2}{\partial
x_i\partial
x_j}\frac{1}{r}=-\frac{4\pi}{3}f(0)\delta_{ij}+\lim_{\varepsilon\to
0+}\int_{r>\varepsilon} {\rm d}^3r\, f(\bold{
r})\,\frac{3x_ix_j-r^2\delta_{ij}}{r^5}.
\end{equation}
Using (38) in (36) results in (35). {\it QED}

The author is grateful to F Farassat for a useful discussion on the
matter of  this appendix.

\end{document}